\documentstyle[12pt,a4,amssymbols,subeqnarray]{article} 

\renewcommand{\theequation}{\thesection.\arabic{equation}} 
\newcommand{\bibiteml}[2]{
                    \bibitem{#1}{#2}
                          }
\newcommand{\calle}[1]{(\ref{#1})}

\begin{document}

\rightline{PLB Ref. No: 0424}
\rightline{TAUP 2308-95 / solv-int 9512006}
\rightline{revised version}

\vspace{1cm}

\begin{center}
{\Large  INTEGRABLE DEFORMATIONS
\\ \vspace{2mm} 
OF THE CFT ON HYPER-ELLIPTIC SURFACES}
\end{center}

\begin{center}
S. A. Apikyan$^\dagger$\\Theoretical Physics Department\\
Yerevan Physics Institute\\ Alikhanyan Br.st.2, Yerevan, 375036 ARMENIA
\end{center}

\vspace{5mm}
\begin{center}
C. J. Efthimiou$^\ddagger$\\
Department of Physics and Astronomy\\Tel-Aviv University\\
Tel Aviv, 69978 ISRAEL
\end{center}

\vspace{2cm}

\begin{abstract}
In this letter, we continue  the work we started at  \cite{AE} and
we propose  new series of integrable models in quantum field theory.
These  models are  obtained as  perturbed models of the minimal 
conformal field theories 
on the hyper-elliptic surfaces by particular relevant operators
of the untwisted sector.
The quantum group symmetry of the models is also discussed.
\end{abstract}

\vfill
\hrule
\ \\
\noindent
$\dagger$ e-mail address: apikyan@vx2.yerphi.am    \\
$\ddagger$ e-mail address: costas@ccsg.tau.ac.il 
\newpage

\section{Introduction}

During the last years the direction of integrable models in
mathematical physics has exprerienced a rapid development.
In particular, integrable theories from Conformal Field Theory
(CFT) have been extremely succesfull due to the revolutionary
ideas of A. Zamolodchikov \cite{Mus}.
However, despite of all developments in this area, our knowledge
on integrable models cannot be considered as complete as there are
still many open problems. For example, the calculation of correlation
functions for integrable models is a problem not completely solved
up this to this day.
 Moreover, almost all known integrable models have been constructed
on the surface of the sphere; the study of integrable models on
higher genus surfaces, although very exciting, has not
progressed very much.

Today Riemann surfaces have invaded physics.
Among them, the category  of  hyper-elliptic surfaces (HESs) is
very special, not only for its simplictity, but also for the
fact that it has appeared in many places in theoretical
physics. For example, HESs appear in string theory \cite{K}, in the 
Ashkin-Teller model of statistical mechanics \cite{zam87} etc.
  One of the most unexpected results was their appearence
in the Seiberg-Witten theory of electric-magnetic duality \cite{AF}.

For the reasons mentioned above, it is therefore quite favorable to explore various
aspects and constuctions on  HESs.
In particular, in this letter,
we continue our  investigation \cite{AE} for constructing
integrable models on HESs.

\section{CFT on HESs}

In this short section we review briefly the main aspects of the 
CFT\footnote{We should mention here that Krichever and Novikov have
also found generalizations
of the Virasoro algebra on higher genus surfaces; their work has been discussed and
extended by various
other authors \cite{KN}. Notice that among Riemann surfaces, the 
HESs are special in the sense that they possess additional ${\Bbb Z}_2$ symmetry
not present in all surfaces.} on HESs.
For a longer discussion, we refer our reader to 
 \cite{CSS} 
or to section 2 of \cite{AE}.  

A HES  $\Gamma$ is a compact Riemann surface of genus $g\geq 1$ determined
by an equation of the form
\begin{equation}
\label{form2}
y^2=P_{2g+2}(z)~,
\end{equation}
where $P_{2g+2}(z)$ is a polynomial of degree $2g+2$. $\Gamma$ is therefore a
double cover of the sphere and every field $\Phi$ on the sphere, under the
hyper-elliptic map \calle{form2}, maps into two fields $\Phi^{(l)},~l=0,1$
living on the two sheets of $\Gamma$.

If $A_a,\, B_a,~a=1,2,\dots,g$ denote the basic cycles of the HES,
the monodromy operators $\hat{\pi}_{A_a}$, $\hat{\pi}_{B_a}$ form a
a representation of the ${\Bbb Z}_2$ group. Therefore, a CFT built on
the HESs will have two sectors: one untwisted (meaning trivial
monodromies) and one twisted (meaning non-trivial monodromies).
In particular, for the energy momentum tensor in the diagonal
basis $T, T^-$, we have
\begin{subeqnarray}
\hat{\pi}_{A_{a}}T=T~,&& \quad \hat{\pi}_{A_{a}}T^{-}=T^{-}~,
\label{BC1}\\
\hat{\pi}_{B_{a}}T=T~,&& \quad \hat{\pi}_{B_{a}}T^{-}=T^{-}~,
\label{BC2}
\end{subeqnarray}
in the untwisted sector and
\begin{subeqnarray}
\hat{\pi}_{A_{a}}T=T~,&& \quad \hat{\pi}_{A_{a}}T^{-}=T^{-}~,
\label{BC3}\\
\hat{\pi}_{B_{a}}T=T~,&& \quad \hat{\pi}_{B_{a}}T^{-}=-T^{-}~,
\label{BC4}
\end{subeqnarray}
in the twisted sector.
The corresponding operator product expansions are
\begin{subeqnarray}
T(z_{1})T(z_{2})&=&{c/2\over z_{12}^4}+
{2\,T(z_2)\over z_{12}^2}+
{T'(z_2)\over z_{12}} + {\rm reg}~,
\label{OPE1} \\
T^{-}(z_1)T^{-}(z_{2})&=&{c/2\over z_{12}^4}+
{2\,T(z_2)\over z_{12}^2}+
{T'(z_2)\over z_{12}} + {\rm reg}~,
\label{OPE2}\\
T(z_1)T^{-}(z_2)&=&{2\over z_{12}^2}\,T^{-}(z_2)+
{T'^{-}(z_2)\over z_{12}}+ {\rm reg}~.
\label{OPE3}
\end{subeqnarray}
 To write the algebra
\calle{OPE1}-\calle{OPE2} in
the graded form we determine the mode expansion of $T$ and $T^-$:
\begin{subeqnarray}
T(z)V_{(k)}(0)&=&\sum_{n\in {\Bbb Z}}\, z^{n-2}L_{-n}V_{(k)}(0)~,\\
T^-(z)V_{(k)}(0)&=&\sum_{n\in {\Bbb Z}}\, z^{n-2-k/2}L_{-n+k/2}^-V_{(k)}(0)~,
\end{subeqnarray}
where $k$ ranges over the values 0,1 and determines the parity sector.

In a Coulomb Gas Formalism, the operators $T$ , $T^-$ are given by
\begin{subeqnarray}
T &=& -\frac{1}{4}(\partial_z\Phi)^2 + i\alpha_0\partial_z^2\Phi -
\frac{1}{4}(\partial_z \Phi^-)^2~, \\
T^- &=& -\frac{1}{2}\partial_z\Phi\partial_z\Phi^- +
i\alpha_0\partial_z^2\Phi^- ~,
\end{subeqnarray}
where the fields $\Phi,\Phi^-$ are two free bosons.
In analogy with the monodromies of the $T,T^-$ fields,
in the untwisted
sector the bosons have trivial monodromies
\begin{subeqnarray}
\hat{\pi}_{A_a}\partial_z\Phi = \partial_z\Phi~,  &&
\hat{\pi}_{B_a}\partial_z\Phi = \partial_z\Phi~,\\
\hat{\pi}_{A_a}\partial_z\Phi^- = \partial_z\Phi^- ~, &&
\hat{\pi}_{B_a}\partial_z\Phi^- = \partial_z\Phi^-~,
\end{subeqnarray}
while in the twisted sector one of them acquires a minus sign:
\begin{subeqnarray}
\hat{\pi}_{A_a}\partial_z\Phi = \partial_z\Phi~,  &&
\hat{\pi}_{B_a}\partial_z\Phi = \partial_z\Phi~,\\
\hat{\pi}_{A_a}\partial_z\Phi^- = \partial_z\Phi^- ~, &&
\hat{\pi}_{B_a}\partial_z\Phi^- = -\partial_z\Phi^-~.
\slabel{minusmonodromy}
\end{subeqnarray}

In the $k=0$ (untwisted) sector, the vertex operator
 with charges $\alpha$, $\beta$
 is given by
\begin{equation}
\label{vertex1}
V_{\alpha\beta}(z) = e^{i\alpha\Phi+i\beta\Phi^-}~,
\end{equation}
with conformal weights  $\Delta=\alpha^2-2\alpha_0\alpha
+\beta^2$ and $\Delta^-=2\alpha\beta-2\alpha_0\beta$.

In the $k=1$ (twisted) sector 
there
is a twist field $\sigma_\epsilon,~\epsilon=0,1$ for which
\begin{equation}
 i\partial_z\Phi^-(z)\sigma_{\epsilon}(0)=
\frac{1}{2}z^{-1/2}\hat{\sigma}_{\epsilon}(0) +
\ldots~.
\end{equation}
Then, in this sector,
the  primary
fields can be written as 
\begin{equation}
\label{vertex2}
V_{\gamma\,\epsilon}^{(t)}=e^{i\gamma\Phi}\sigma_{\epsilon}~ ,
\end{equation}
and have  weight
$\Delta^{(t)}=\gamma^2-2\alpha_0\gamma+1/16.$

For the unitary minimal models on the HES the central charge takes 
the form
\begin{equation}
  c=2-{12\over p(p+1)}~,~~~~~p=3,4,\dots~.
\end{equation}
The corresponding values of $\alpha, \beta, \gamma$ charges are
\begin{equation}
\begin{array}{l}
\alpha_{n'm'}^{nm}={2-n-n'\over 2}\,\alpha_{+}+
{2-m-m'\over 2}\,\alpha_{-}~,\\
\nonumber\\
\beta_{n'm'}^{nm}={n-n'\over 2}\,\alpha_{+}+
{m-m'\over 2}\,\alpha_{-}~,\\
\nonumber\\
\gamma_{nm}={2-n\over 2}\,\alpha_{+}+
{2-m\over 2}\,\alpha_{-}~,\\
\nonumber\\
1\leq n,n'\leq p ,\quad 1\leq m,m'\leq p-1~,
\end{array}
\end{equation}
where the constants $\alpha_{\pm}$
are expressed in terms of the background charge $\alpha_0$:
\begin{equation}
\alpha_{\pm}=\alpha_{0}/2 \pm \sqrt{\alpha_{0}^{2}/4+1/2} ~.
\end{equation}
We denote the corresponding fields by $V^{nm}_{n'm'}$, $V^{(t)}_{nm}$
and their conformal weights by $\Delta^{nm}_{n'm'}$, $\Delta^{(t)}_{nm}$.
Explicitly
\begin{subeqnarray}
 \Delta^{nm}_{n'm'} &=& {  [(n+n')(p+1)-(m+m')p]^2 +
                          [(n-n')(p+1)-(m-m')p]^2 -4\over
                          8p(p+1) }~,~~~~~~~~~\\
 \Delta^{(t)}_{nm} &=& {  [(p+1)n-mp]^2-4\over 8p(p+1) }
                       +{1\over 16}~.
\end{subeqnarray}

\section{Integrable Deformations of the CFT on the HESs} 
\setcounter{equation}{0}

In \cite{AE}, we investigated the integrability of models costructed 
as deformations of the unitary minimal models
$S_p\lbrack\Phi,\Phi^-\rbrack$ on HES by primary fields in the {\em twisted}
sector. In particular, we found that among all series of  models
\begin{equation}
S_\lambda\lbrack n\,m;p\rbrack\,=\,S_p\lbrack\Phi,\Phi^-\rbrack
+{\lambda\over 2\pi}\,\int\,d^2z\, V^{(t)}_{nm}~,
\end{equation}
only the series perturbed by $V^{(t)}_{11}$ is integrable:
\begin{equation}
S_\lambda\lbrack 1~1;p\rbrack ~=~ S_p\lbrack\Phi,\Phi^-\rbrack
+{\lambda\over 2\pi}\,\int\,d^2z\, 
e^{i\gamma_{11}
\Phi(z,\overline{z})}\,\sigma_{\epsilon}(z,\overline{z})~.
\end{equation}
The models beloging in this series are all  massive 
and there is no indication of non-trivial
fixed points.

Now, we examine deformations of the unitary minimal models on HESs
by primary fields in the {\em untwisted} sector, i.e. we seek
integrable models of the form:
\begin{equation}
S_{\lambda}\left\lbrack\matrix{n&m\cr n'&m'\cr}\, ; \, p \right\rbrack
 ~=~ S_p\lbrack\Phi,\Phi^{-}\rbrack\,+
\,{\lambda\over 2\pi}\,\int\,d^2z\,
V_{n'm'}^{nm}(z,\overline z)~.
\end{equation}
The parameter $\lambda$ is a coupling constant
with  conformal weights
$(1-\Delta^{nm}_{n'm'},1-\Delta^{nm}_{n'm'})$ with respect to
$T$.

For a generic perturbation 
$S_{\lambda}\left\lbrack\matrix{n&m\cr n'&m'\cr}\, ;\,p\right\rbrack$
cannot be integrable; however, one may be able to choose the primary field
$V_{n'm'}^{nm}(z,\overline z)$ such that it becomes integrable.
To this end, we search for relevant operators for which 
Zamolochikov's counting argument \cite{zam89}
guarantees the existence of conserved
charges. 
It is not very hard to see that 
the  following models
\begin{subeqnarray}
\label{mnaction}
S_{\lambda}\left\lbrack\matrix{1&1\cr n&m\cr}\, ;\, p \right\rbrack &=&
S_p\lbrack\Phi,\Phi^-\rbrack\,+
\,{\lambda\over 2\pi}\,\int\,d^2z \, V^{1,1}_{nm}(z,\overline z)~,
\end{subeqnarray}
where
\begin{equation}
  m= n-1, ~n,~n+1, ~n+2~.
\end{equation}
are almost all integrable.
Some comments are in order here:
\begin{itemize}
\item 
  `Almost all models' means that, for some values of $n,m$, there is
  a minimum number $p_{\min}(m,n)$ such that all models 
  with $p>p_{\min}$ are integrable. When $p\le p_{\min}$,
  the perturbations are not relevant.
\item
   As usual, integrability may be spoiled by resonance conditions.
   However, the phenomenon of resonance conditions is not generic 
   and we will simply interpret the result as
   `integrability modulo resonance conditions'.
\item
  Notice that the fields $V_{11}^{nm}, V_{nm}^{11}$ have the
  same  weights relative
  to $T$ (and opposite weight relative to $T^-$). 
  Zamolodchikov's
  counting argument the arguments is only sensitive to the
  weights relative to $T$.
\end{itemize}

In the following, we shall concentrate (for the shake of presentation)
on the  model with $n=1,~m=3$; 
a similar discussion for the rest models can be also given
with few changes.
The beta function $\beta(g)$
of the renormalized coupling constant $g=g(\lambda)$
for the first model is given by the equation
\begin{equation}
\beta_g=\varepsilon \,g-{C\over 2}\, g^2 +{\cal O}(g^3) ~,
 \quad \varepsilon=1-\Delta^{1,1}_{1,3}={2\over p+1}~,
\end{equation}
where $C$ is the structure constant
\begin{equation}
 \langle V^{1,1}_{1,3}(z_1,\overline z_1) 
 V^{1,1}_{1,3}(z_2,\overline z_2) V^{1,1}_{1,3}(z_3,\overline z_3)\rangle ~=~ 
  {C\over |z_{12}|^{2\Delta^{1,1}_{1,3}}
          |z_{23}|^{2\Delta^{1,1}_{1,3}}
          |z_{31}|^{2\Delta^{1,1}_{1,3}}   } ~.
\end{equation}
Depending
on the sign of $\lambda$, we  have either a  massive or
a massless theory. In particular, for $\lambda <0$ we do not expect any
fixed points and therefore the model is massive; for $\lambda>0$ there is
a fixed point and therefore the model flows to a massless theory.
We can make this statement more exact by considering large values of
$p$. Then one finds
\begin{equation}
   \beta_g ~=~
             g\,\varepsilon- {2\over\sqrt{3}}\, \left(1-{3\varepsilon\over2}
          \right)\, g^2 
   + {\cal O}(g^3) ~.
\end{equation}
When the coupling constant is negative, there is a non-trivial fixed point
\begin{equation}
   g_* ~=~ {\sqrt{3}\over 2}\,\varepsilon\,
  \left( 1+{3\varepsilon\over2}
  \right) + {\cal O}(\varepsilon^3)~,
\end{equation}
which corresponds to the
$S_{p-1}\lbrack \Phi,\Phi^-\rbrack$ unitary minimal model of
CFT on HES; therefore, in the infrared the theory becomes massless.

\section{The Quantum Group Symmetry of the Models}
\setcounter{equation}{0}

The method  of non-local conserved charges \cite{BL}
 has been proposed as an alternative
to the inverse quantum scattering method. The basic ingredient of the
method is the construction of a maximum number of conserved currents
\begin{subeqnarray}
\partial_{\overline{z}}J &=& \partial_{z}H ~,\\
\partial_{z}\overline{J} &=& \partial_{\overline{z}}
\overline{H}~,
\end{subeqnarray}
which obey non-trivial braiding relations. These laws lead
to the conserved charges
\begin{subeqnarray}
 Q &=&\int\, {dz\over 2\pi i}\,J +\int\,{d\overline z\over 2\pi i}\,  H ~,\\
 \overline Q &=&\int\, {d\overline z\over 2\pi i}\,\overline J +
  \int\,{dz\over 2\pi i} \,  
 \overline H ~,
\end{subeqnarray}
or more compactly
\begin{equation}
    Q
   =\int_{\cal C}\, dz^\mu\,\epsilon_{\mu\nu} J^\nu(z)~,
\end{equation}
which act on the Hilbert space ${\cal H}=\{ \Phi^k\}$
$$
   Q(\Phi^k(w))
   =\int_{{\cal C}(w)}\, dz^\mu\,\epsilon_{\mu\nu} J^\nu(z)\Phi^k(w)~.
$$
with the following subtlety taken into account: the charge $Q$ is
{\em not}  defined by integrating the currents along an equal time slice,
but along a path ${\cal C}(w)$ from $-\infty$ to $-\infty$ surrounding the
point $w$ of the field and avoiding the cuts on the $z$ plane.

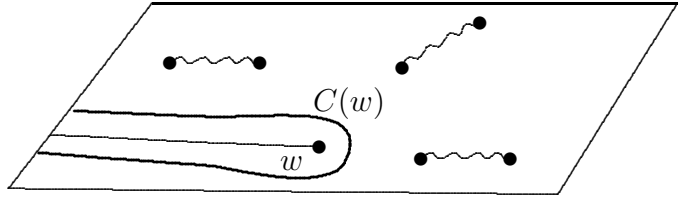
\input prepictex
\input pictex
\input postpictex

\centerline{
\font\thinlinefont=cmr5
\begingroup\makeatletter\ifx\SetFigFont\undefined
\def\x#1#2#3#4#5#6#7\relax{\def\x{#1#2#3#4#5#6}}%
\expandafter\x\fmtname xxxxxx\relax \def\y{splain}%
\ifx\x\y   
\gdef\SetFigFont#1#2#3{%
  \ifnum #1<17\tiny\else \ifnum #1<20\small\else
  \ifnum #1<24\normalsize\else \ifnum #1<29\large\else
  \ifnum #1<34\Large\else \ifnum #1<41\LARGE\else
     \huge\fi\fi\fi\fi\fi\fi
  \csname #3\endcsname}%
\else
\gdef\SetFigFont#1#2#3{\begingroup
  \count@#1\relax \ifnum 25<\count@\count@25\fi
  \def\x{\endgroup\@setsize\SetFigFont{#2pt}}%
  \expandafter\x
    \csname \romannumeral\the\count@ pt\expandafter\endcsname
    \csname @\romannumeral\the\count@ pt\endcsname
  \csname #3\endcsname}%
\fi
\fi\endgroup
\mbox{\beginpicture
\setcoordinatesystem units <.5cm,.5cm>
\unitlength=.5cm
\linethickness=1pt
\setplotsymbol ({\makebox(0,0)[l]{\tencirc\symbol{'160}}})
\setshadesymbol ({\thinlinefont .})
\setlinear
\linethickness= 0.500pt
\setplotsymbol ({\thinlinefont .})
%
%
\plot  7.463 19.365 	 7.564 19.448
	 7.646 19.500
	 7.716 19.524
	 7.781 19.524
	 7.862 19.446
	 7.940 19.365
	 8.057 19.380
	 8.178 19.445
	 8.299 19.509
	 8.416 19.524
	 8.495 19.445
	 8.575 19.365
	 8.692 19.380
	 8.813 19.445
	 8.934 19.509
	 9.051 19.524
	 9.128 19.443
	 9.210 19.365
	 9.292 19.385
	 9.368 19.445
	 9.445 19.505
	 9.527 19.524
	 9.603 19.483
	 9.686 19.365
	/
%
%
\linethickness= 0.500pt
\setplotsymbol ({\thinlinefont .})
\put{\makebox(0,0)[l]{\circle*{ 0.318}}} at 11.430 17.145
%
%
\linethickness= 0.500pt
\setplotsymbol ({\thinlinefont .})
\putrule from  6.985 20.955 to 20.955 20.955
\plot 20.955 20.955 17.780 15.875 /
\plot 17.780 15.875  3.175 16.034 /
\plot  3.175 16.034  6.985 20.955 /
\linethickness= 0.500pt
\setplotsymbol ({\thinlinefont .})
%
%
\plot  4.286 17.462 	 4.383 17.458
	 4.478 17.453
	 4.573 17.448
	 4.667 17.443
	 4.760 17.438
	 4.853 17.434
	 4.944 17.429
	 5.035 17.424
	 5.124 17.420
	 5.213 17.415
	 5.302 17.411
	 5.389 17.407
	 5.475 17.402
	 5.561 17.398
	 5.646 17.394
	 5.730 17.389
	 5.814 17.385
	 5.897 17.381
	 5.978 17.377
	 6.060 17.373
	 6.140 17.369
	 6.220 17.365
	 6.298 17.361
	 6.376 17.357
	 6.454 17.353
	 6.531 17.350
	 6.606 17.346
	 6.682 17.342
	 6.756 17.338
	 6.830 17.335
	 6.903 17.331
	 6.975 17.328
	 7.047 17.324
	 7.118 17.321
	 7.188 17.317
	 7.258 17.314
	 7.327 17.311
	 7.395 17.307
	 7.463 17.304
	 7.530 17.301
	 7.596 17.298
	 7.661 17.295
	 7.726 17.291
	 7.791 17.288
	 7.917 17.282
	 8.042 17.276
	 8.164 17.271
	 8.283 17.265
	 8.400 17.260
	 8.515 17.255
	 8.627 17.249
	 8.737 17.244
	 8.845 17.240
	 8.951 17.235
	 9.054 17.230
	 9.156 17.226
	 9.255 17.221
	 9.353 17.217
	 9.448 17.213
	 9.542 17.209
	 9.633 17.205
	 9.723 17.202
	 9.811 17.198
	 9.897 17.195
	 9.982 17.191
	10.065 17.188
	10.146 17.185
	10.225 17.182
	10.303 17.179
	10.380 17.176
	10.455 17.174
	10.528 17.171
	10.600 17.169
	10.671 17.166
	10.741 17.164
	10.809 17.162
	10.876 17.160
	10.942 17.158
	11.006 17.156
	11.070 17.154
	11.194 17.151
	11.314 17.148
	11.430 17.145
	11.430 17.145
	/
\linethickness=1pt
\setplotsymbol ({\makebox(0,0)[l]{\tencirc\symbol{'160}}})
%
%
\plot  4.921 18.098 	 5.019 18.093
	 5.116 18.089
	 5.211 18.085
	 5.304 18.081
	 5.395 18.077
	 5.485 18.073
	 5.573 18.070
	 5.660 18.066
	 5.745 18.062
	 5.828 18.059
	 5.910 18.055
	 5.990 18.052
	 6.069 18.048
	 6.146 18.045
	 6.222 18.042
	 6.296 18.038
	 6.369 18.035
	 6.441 18.032
	 6.511 18.029
	 6.579 18.026
	 6.647 18.023
	 6.713 18.020
	 6.778 18.018
	 6.841 18.015
	 6.965 18.010
	 7.083 18.004
	 7.197 18.000
	 7.306 17.995
	 7.411 17.990
	 7.511 17.986
	 7.608 17.982
	 7.701 17.978
	 7.790 17.974
	 7.875 17.970
	 7.957 17.966
	 8.036 17.963
	 8.112 17.959
	 8.185 17.956
	 8.255 17.953
	 8.323 17.950
	 8.389 17.947
	 8.452 17.944
	 8.572 17.939
	 8.647 17.936
	 8.726 17.935
	 8.811 17.935
	 8.901 17.937
	 8.995 17.939
	 9.093 17.942
	 9.195 17.946
	 9.300 17.950
	 9.408 17.954
	 9.519 17.959
	 9.633 17.963
	 9.748 17.968
	 9.865 17.972
	 9.983 17.975
	10.102 17.978
	10.221 17.980
	10.341 17.982
	10.460 17.982
	10.579 17.981
	10.697 17.978
	10.813 17.974
	10.928 17.968
	11.041 17.961
	11.151 17.951
	11.259 17.939
	11.364 17.924
	11.465 17.908
	11.562 17.888
	11.655 17.866
	11.744 17.840
	11.828 17.812
	11.906 17.780
	11.995 17.726
	12.088 17.645
	12.170 17.553
	12.224 17.462
	12.249 17.354
	12.254 17.290
	12.255 17.223
	12.252 17.156
	12.245 17.092
	12.224 16.986
	12.178 16.865
	12.145 16.795
	12.107 16.725
	12.063 16.658
	12.014 16.598
	11.906 16.510
	11.813 16.467
	11.714 16.430
	11.611 16.399
	11.504 16.373
	11.393 16.353
	11.279 16.337
	11.162 16.326
	11.042 16.320
	10.920 16.317
	10.796 16.318
	10.670 16.323
	10.606 16.326
	10.542 16.330
	10.478 16.335
	10.414 16.341
	10.350 16.347
	10.286 16.354
	10.221 16.361
	10.156 16.369
	10.092 16.377
	10.027 16.386
	 9.963 16.395
	 9.899 16.404
	 9.835 16.414
	 9.771 16.424
	 9.644 16.445
	 9.519 16.467
	 9.396 16.489
	 9.274 16.511
	 9.155 16.533
	 9.039 16.554
	 8.927 16.574
	 8.817 16.594
	 8.712 16.612
	 8.611 16.628
	 8.514 16.642
	 8.422 16.653
	 8.336 16.663
	 8.255 16.669
	 8.185 16.673
	 8.113 16.677
	 8.039 16.682
	 7.962 16.687
	 7.882 16.692
	 7.800 16.697
	 7.714 16.703
	 7.625 16.709
	 7.532 16.715
	 7.436 16.722
	 7.335 16.729
	 7.231 16.737
	 7.122 16.745
	 7.009 16.753
	 6.890 16.761
	 6.767 16.771
	 6.704 16.775
	 6.639 16.780
	 6.573 16.785
	 6.506 16.790
	 6.437 16.795
	 6.367 16.800
	 6.295 16.806
	 6.222 16.811
	 6.148 16.817
	 6.071 16.823
	 5.994 16.829
	 5.915 16.835
	 5.834 16.841
	 5.752 16.847
	 5.668 16.854
	 5.582 16.860
	 5.495 16.867
	 5.406 16.874
	 5.315 16.881
	 5.223 16.888
	 5.129 16.896
	 5.033 16.903
	 4.935 16.911
	 4.835 16.918
	 4.734 16.926
	 4.630 16.934
	 4.525 16.943
	 4.418 16.951
	 4.309 16.960
	 4.197 16.968
	 4.084 16.977
	 3.969 16.986
	/
%
%
\put{$C(w)$} at 12.224 18.3
%
%
\put{$w$}  at 10.7 16.7
%
%
\linethickness= 0.500pt
\setplotsymbol ({\thinlinefont .})
\put{\makebox(0,0)[l]{\circle*{ 0.318}}} at  9.845 19.365
%
%
\linethickness= 0.500pt
\setplotsymbol ({\thinlinefont .})
\put{\makebox(0,0)[l]{\circle*{ 0.318}}} at 13.636 19.243
%
%
\linethickness= 0.500pt
\setplotsymbol ({\thinlinefont .})
\put{\makebox(0,0)[l]{\circle*{ 0.318}}} at 15.699 20.434
\linethickness= 0.500pt
\setplotsymbol ({\thinlinefont .})
%
%
\plot 13.636 19.243 	13.682 19.365
	13.727 19.451
	13.832 19.539
	13.941 19.513
	14.048 19.482
	14.142 19.553
	14.216 19.669
	14.289 19.785
	14.383 19.856
	14.491 19.828
	14.599 19.799
	14.692 19.870
	14.765 19.987
	14.837 20.103
	14.931 20.174
	15.039 20.143
	15.149 20.117
	15.210 20.174
	15.246 20.264
	15.283 20.353
	15.344 20.411
	15.431 20.413
	15.488 20.392
	15.562 20.354
	/
%
%
\linethickness= 0.500pt
\setplotsymbol ({\thinlinefont .})
\put{\makebox(0,0)[l]{\circle*{ 0.318}}} at 14.131 16.825
%
%
\linethickness= 0.500pt
\setplotsymbol ({\thinlinefont .})
\put{\makebox(0,0)[l]{\circle*{ 0.318}}} at 16.512 16.825
\linethickness= 0.500pt
\setplotsymbol ({\thinlinefont .})
%
%
\plot 14.131 16.825 	14.232 16.908
	14.314 16.960
	14.383 16.984
	14.448 16.984
	14.530 16.906
	14.607 16.825
	14.724 16.840
	14.845 16.905
	14.966 16.969
	15.083 16.984
	15.163 16.905
	15.242 16.825
	15.359 16.840
	15.480 16.905
	15.601 16.969
	15.718 16.984
	15.796 16.903
	15.877 16.825
	15.959 16.845
	16.036 16.905
	16.112 16.965
	16.195 16.984
	16.271 16.943
	16.353 16.825
	/
%
%
\linethickness= 0.500pt
\setplotsymbol ({\thinlinefont .})
\put{\makebox(0,0)[l]{\circle*{ 0.318}}} at  7.463 19.365
\linethickness=0pt
\putrectangle corners at  3.150 20.980 and 20.980 15.850
\endpicture}
}

Here, we shall not discuss the way the Hilbert
space can be constructed. We point out that such a 
method of construction
has been given in \cite{BL,FL,E} and we refer the interested
reader to these papers.
The conserved charges constructed as above  generate a quantum group which, 
in turn, determines\footnote{Notice that all
quantities related with the scattering of particles should be
considered on  the punctured sphere; in this case
 the usual LSZ method \cite{Ryder}, which requires well defined asymptotic
states for the calculation of the S-matrix,
is meaningful.}
the S-matrix of the model.

For the action $S_\lambda[^{11}_{13}]$,
one has the following  non-local currents:
\begin{subeqnarray}
\label{currents}
J_{\pm}(z) &=& e^{\pm i{1\over 2\alpha_-}\,(\varphi(z)+\varphi^-(z))}~, \\
H_{\pm}(z,\bar{z}) &=& \lambda \frac{2\alpha_-^2}{2\alpha_-^2 - 1}\,
e^{\pm i({1\over 2\alpha_-} -\alpha_-)(\varphi(z)+\varphi^-(z))
\mp i\alpha_-\, (\bar{\varphi}(\bar{z})+\bar{\varphi}^-(\bar{z}))}
 ~, \\
\overline{J}_{\pm}(\bar{z}) &=& e^{\mp i{1\over 2\alpha_-}\,
(\bar{\varphi}(\bar{z})+
\bar{\varphi}^-(\bar{z}))}
  ~, \\
\overline{H}_{\pm}(z,\bar{z}) &=& \lambda \frac{2\alpha_-^2}{2\alpha_-^2 - 1}
e^{\mp i({1\over 2\alpha_-} -\alpha_-)\,
(\bar{\varphi}(\bar{z})+\bar{\varphi}^-(\bar{z}))
\pm i\alpha_-\, (\varphi(z) +\varphi^-(z))}    ~.
\end{subeqnarray}

The conserved charges arising from the above currents 
satisfy the following commutation relations:
\begin{subeqnarray}
 \lbrack Q_+,\overline Q_+\rbrack_{q^{\phantom{-1}}} &=&
 \lbrack Q_-,\overline Q_-\rbrack_q ~=~ 0~,\\
 \lbrack Q_+, \overline Q_-\rbrack_{q^{-1}} &=& a\, (1-q^{2T})~,\\
 \lbrack Q_-, \overline Q_+\rbrack_{q^{-1}} &=& a\, (1-q^{-2T})~,
\end{subeqnarray}
where 
\begin{equation}
 \lbrack A, B\rbrack_q ~\equiv~ A\, B - q^2\, B\, A~,
\end{equation}
and
\begin{equation}
q=e^{-i \pi/2\alpha_-^2}~,~~~~~a={\lambda\over 2\pi i}\,
 ({1\over 2\alpha_-^2}-1)~.
\end{equation}
Finally, the topological charge is defined by
\begin{equation}
   T~=~ {\alpha_-\over 2\pi}\, \int\,dx \, \partial_x(\Phi+\Phi^-)~.
\end{equation}
                           
The algebra defined by the charges $Q_\pm,\overline Q_\pm, T$ is a known
infinite dimensional algebra, namely the the quantum deformation
of the ${\frak sl}\,(2)$ affine Kac-Moody algebra with vanishing
center; the symbol $\widehat{{\frak sl}_{\,q}(2)}$ is usually
 used to denote this
algebra.

Requiring the S-matrix to commute with the
non-local charges, we obtain very stringent constraints
\begin{equation}
     \lbrack S, \Delta(Q_\pm)\rbrack ~=~
     \lbrack S, \Delta(\overline Q_\pm)\rbrack ~=~
     \lbrack S, \Delta(T)\rbrack ~=~ 0~,
\end{equation}
that determine
the S-matrix up to a scalar factor
\cite{BL}.
In the above equations, $\Delta(Q)$ denotes the action of the charge
$Q$ on an asymptotic pair of solitons.

Before closing the section, we point out that when $n=1, m>1$
the qunatum group structure is similar to the one described above,
while in the case $m,n>1$ we have two copies of 
$\widehat{{\frak sl}_q\,(2)}$.

\section{Discussion}

In this letter we have constructed  new integrable models in quantum
field theory using deformations of the CFT on
HESs by relevant operators.
In fact, the CFT on HESs is
equivalent to the $WD_2$ algebra. Integrable perturbations of the
$WD_n$ algebras by the relevant operator of scaling dimension $d=\Delta+
\overline \Delta$, where
$$
  \Delta ~=~ 1-{2\,(n-1)\over p\,(p+1)}~,
$$
have been studied by Lukyanov and Fateev \cite{LF}. Moreover, integrable
perturbations of the $W$ algebras that generalize the $\Phi_{12}$-perturbed
Virasoro models have been considered by Vaysburd and  Babichenko \cite{VB,B}.
In the above papers the assumption of a simple Lie algebra has been used.
However, the underlying algebra of $WD_2$ is not simple. In this sense,
our discussion completes the discussion  of the perturbed $W$ algebras
considered by all of the above authors.

\vspace{.5cm}
{\bf Note Added in Proof}

Similar results as the ones reported in section 3 have been obtained
independently by I. Vaysburd \cite{V2}.

\end{document}